\documentclass[journal]{IEEEtran}
\IEEEoverridecommandlockouts
\usepackage{graphicx, amsmath, amssymb, subcaption, algorithm, algorithmic, url, cite, xcolor}
\usepackage[font={small}]{caption}
\usepackage[hidelinks]{hyperref}
\graphicspath{ {./figures/} }

\begin{document}

\title{Optimal Power Allocation for Rate Splitting Communications with Deep Reinforcement Learning}

\author{Nguyen Quang Hieu, Dinh Thai Hoang, \IEEEmembership{Member, IEEE}, Dusit Niyato, \IEEEmembership{Fellow, IEEE}, and Dong In Kim, \IEEEmembership{Fellow, IEEE}
\thanks{
This research is supported, in part, by the National Research Foundation (NRF), Singapore, funded under Energy Research Test-Bed and Industry Partnership Funding Initiative, part of the Energy Grid (EG) 2.0 programme, Alibaba Group through Alibaba Innovative Research (AIR) Program and Alibaba-NTU Singapore Joint Research Institute (JRI), the National Research Foundation, Singapore under the AI Singapore Programme (AISG) (AISG2-RP-2020-019), WASP/NTU grant M4082187 (4080) and Singapore Ministry of Education (MOE) Tier 1 (RG16/20).
}
\thanks{N. Q. Hieu and D. Niyato are with the School of Computer Science and Engineering, Nanyang Technological University, Sinapore  (e-mail: \{quanghieu.nguyen, dniyato\}@ntu.edu.sg. D. T. Hoang is with the School of Electical and Data Engineerng, University of Technology Sydney, Sydney, NSW 2007, Australia (e-mail: hoang.dinh@uts.edu.au). D. I. Kim is with the Department of Electrical and Computer Engineering, Sungkyunkwan University (SKKU), Suwon 16419, South Korea (e-mail: dikim@skku.ac.kr).}
}

\maketitle

\begin{abstract}
This letter introduces a novel framework to optimize the power allocation for users in a Rate Splitting Multiple Access (RSMA) network. In the network, messages intended for users are split into different parts that are a single common part and respective private parts. This mechanism enables RSMA to flexibly manage interference and thus enhance energy and spectral efficiency. Although possessing outstanding advantages, optimizing power allocation in RSMA is very challenging under the uncertainty of the communication channel and the transmitter has limited knowledge of the channel information. To solve the problem, we first develop a Markov Decision Process framework to model the dynamic of the communication channel. The deep reinforcement algorithm is then proposed to find the optimal power allocation policy for the transmitter without requiring any prior information of the channel. The simulation results show that the proposed scheme can outperform baseline schemes in terms of average sum-rate under different power and QoS requirements. 
\end{abstract}
\vspace{-0.2cm}
\IEEEpeerreviewmaketitle

\vspace{-0.3cm}
\section{Introduction}

\IEEEPARstart{R}{ate} splitting multiple access (RSMA) has emerged as a promising technology that can achieve robust, high data rate, low latency for 6G networks. 
RSMA is based on a concept of rate splitting in which each message transmitted from the transmitter to a user is split into a common (public) part and a private part~\cite{dizdar2020}. The common parts of the messages are then combined into a single common message and can be encoded with a public shared codebook. The private parts are independently encoded to respective users. 
At each user, the common message is first decoded by using the public shared codebook among the users and transmitter. 
After that, each user reconstructs its original message from the part of its common message and its intended private message with Successive Interference Cancellation (SIC).
In traditional multiple access methods, the rate performance is affected by the multiuser interference when the Channel State Information at the Transmitter (CSIT) is imperfect. 
In contrast, RSMA can flexibly manage interference by allowing the interference to be partially decoded and partially treated as noise.
Thus, RSMA can enhance the spectral efficiency, energy efficiency, and security, compared to those of existing multiple access schemes, i.e., Space Division Multiple Access (SDMA), Non-Orthogonal Multiple Access (NOMA), Orthogonal Multiple Access (OMA) and multicasting~\cite{dizdar2020, dai2016, joudeh2016, li2020, joudeh2016sumrate}. 

Although possessing some outstanding advantages, optimizing the performance of RSMA in terms of spectral and energy efficiency is very challenging. Unlike traditional methods, messages in RSMA are split into different parts and the transmitter has to allocate carefully transmission power for each split message to meet the power and QoS constraints with the imperfect CSIT. The transmitter can only maintain an estimation of the CSIT based on the feedback from the users. For this, the transmitter is usually assumed to have information of the channel distribution in advance~\cite{dai2016, joudeh2016}.
In~\cite{li2020}, a cooperative rate splitting scheme is proposed to enhance secure sum-rate in an RSMA network by utilizing the common message in two purposes, i.e., a desired message and artificial noise.
In~\cite{joudeh2016sumrate}, a precoder design and sum-rate maximization are jointly optimized in which the channel state is allowed to change during the transmission according to some known stationary distributions.
Although aforementioned works can maximize the (secure) sum-rate under partial or imperfect CSIT, either the channel state distribution or channel state matrix is assumed to be known by the transmitter in advance.
However, this assumption might not be practical, especially in environments that have severe interference caused by constant changes of multiple channels between users~\cite{he2019}.  

In this paper, we introduce a framework that enables the transmitter can adaptively select the power allocation policy under the dynamic and  uncertainty of communication channel. 
For this, we first formulate the power allocation problem by using the Markov Decision Process (MDP) framework. We then introduce a highly-effective deep reinforcement learning (DRL) scheme based on Proximal Policy Optimization~\cite{schulman2017} algorithm to find the optimal policy for the transmitter without requiring any information of the channel in advance.
To the best of our knowledge, this is the first approach using DRL to solve the power allocation problem for RSMA networks.
Simulation results show that our proposed scheme can outperform other baseline schemes in terms of sum-rate and QoS.

\vspace{-0.3cm}
\section{System Model}
\label{sec:system-model}
\vspace{-0.1cm}
\begin{figure}
\centering
\includegraphics[width=0.7\linewidth]{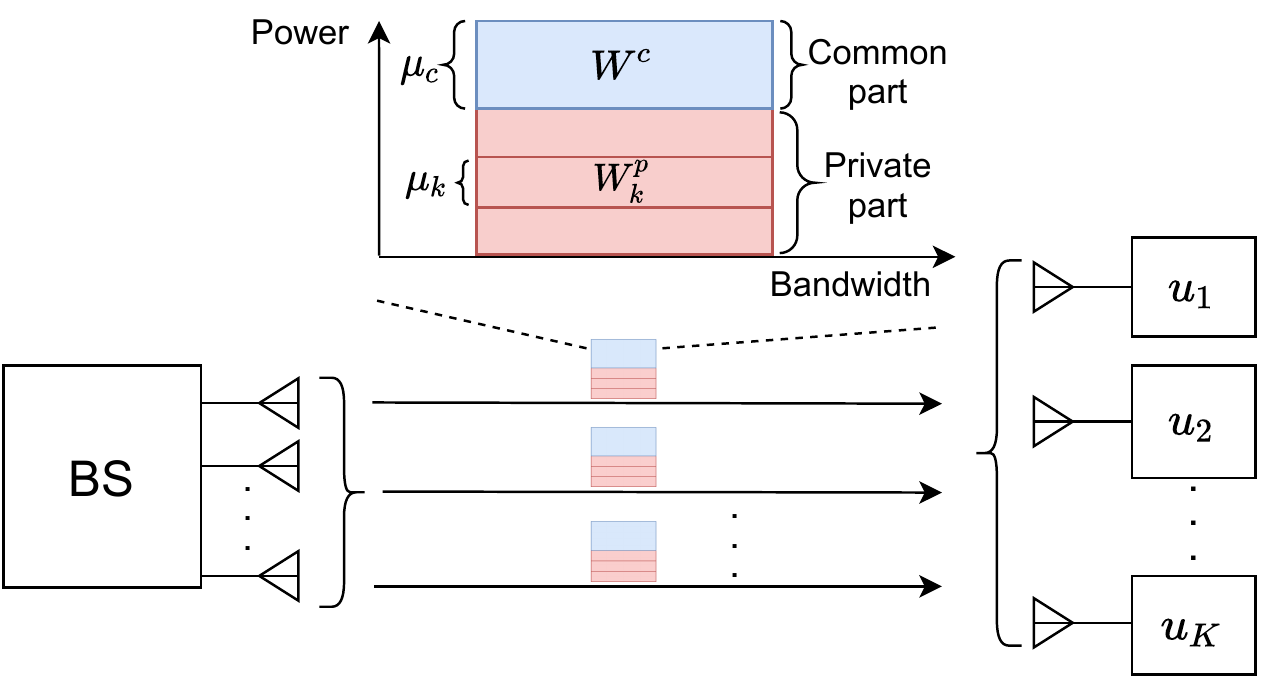}
\caption{An RSMA network consists of $1$ $M$-antenna Base Station (BS) and $K$ single-antenna users.}
\label{fig:system-model}
\vspace{-0.2cm}
\end{figure}

We consider an RSMA network which consists of one $M$-antenna Base Station (BS) and $K$ single-antenna users ($M \geq K$), denoted by $\mathbf{U} = \{u_1, \ldots, u_k, \ldots, u_K\}$, as illustrated in Fig.~\ref{fig:system-model}.
The BS has a set of messages $\mathbf{W} = \{W_1, \ldots, W_k, \ldots, W_K\}$ to be transmitted to the users.
The message intended for user $u_k$, denoted as $W_k$, is split into a common part and a private part, i.e., $W_k^c$ and $W_k^p$ ($\forall k \in \mathcal{K}$ with $\mathcal{K} = \{1, 2, \ldots, K\}$), respectively. The common parts of all $K$ messages are combined into a single common message $W^c$. The single common message $W^c$ and $K$ private messages $W_k^p$ are independently encoded into streams $s_c, s_1, s_2, \ldots, s_K$, where $s_c$ and $s_k$ are encoded common and private symbols. 
The transmitted signal of the BS is thus defined as follows:
\vspace{-0.2cm}
\begin{equation}
\mathbf{x} = \sqrt{\mu_c P_t} \mathbf{w}_c s_c + \sum_{k=1}^{K} \sqrt{\mu_k P_t} \mathbf{w}_k s_k, 
\vspace{-0.1cm}
\end{equation}
where $\mathbf{w}_c \in \mathbb{C}^{M \times 1}$ and $\mathbf{w}_k \in \mathbb{C}^{M \times 1}$ are the precoding vectors of the common and private messages, respectively. 
$\mu_c$ and $\mu_k$ are the power allocation coefficients, i.e., the ratios between the transmission power allocated for the common and private messages to the total transmission power $P_t$, respectively. The normalized power allocation coefficients are constrained by $\mu_c + \sum_{ k = 1}^{K} \mu_k \leq 1$.
The received signal at user $u_k$ is $\mathbf{y}_k = \mathbf{h}_k^H \mathbf{x} + n_k$, where $n_k$ is noise at the user, $\mathbf{h}_k \in \mathbb{C}^{M \times 1}$ is the channel gain between the BS and user $u_k$.
The SINRs of the common and private messages are calculated as follows:
\vspace{-0.2cm}
\begin{equation}
\begin{aligned}
\label{eq:sinr}
 & \gamma_k^c(\pmb{\mu})= \frac{\mu_c P_t |\mathbf{h}_k \mathbf{w}_c|^2}{\sum_{j=1}^{K} \mu_j P_t|\mathbf{h}_k \mathbf{w}_j|^2 + 1},  \\
 & \gamma_k^p(\pmb{\mu}) = \frac{\mu_k P_t |\mathbf{h}_k \mathbf{w}_k|^2}{\sum_{j\neq k} \mu_j P_t|\mathbf{h}_k \mathbf{w}_j|^2 + 1},
\end{aligned}
\vspace{-0.1cm}
\end{equation}
where $\pmb{\mu} = [\mu_c,\mu_1, \mu_2, \ldots, \mu_k, \ldots, \mu_K]$ is the power allocation coefficient vector. The noise power is normalized to one for simplicity. 
With the above SINRs, achievable rates of the private messages are calculated as follows:
\vspace{-0.2cm}
\begin{equation}
\vspace{-0.1cm}
R_k(\pmb{\mu}) = \log_2\big(1+\gamma_k^p(\pmb{\mu})\big), \forall k \in \mathcal{K}.
\vspace{-0.1cm}
\end{equation}
To ensure that the common message can be successfully decoded by all users, the achievable rate of the common message is calculated as follows:
\vspace{-0.2cm}
\begin{equation}
\vspace{-0.1cm}
R_c(\pmb{\mu}) =  \min_{k \in \mathcal{K}} \Big\{\log_2 \big(1+\gamma_k^c(\pmb{\mu})\big)\Big\}.
\vspace{-0.1cm}
\end{equation}
Since $R_c$ is shared between users such that $C_k$ is the user $u_k$'s portion of the common rate $R_c$ with $\sum_{k=1}^{K}C_k \leq R_c$. The total achievable rate of the user $u_k$ is then defined by $R_k^{tot} = C_k + R_k$~\cite{yijie2018}. 
The sum-rate is calculated by~\cite{dai2016}: 
\vspace{-0.3cm}
\begin{equation}
R_{sum}(\pmb{\mu, c}) = \sum_{k=1}^{K} \big(C_k(\pmb{\mu}, \mathbf{c}) + R_k(\pmb{\mu})\big),
\vspace{-0.2cm}
\end{equation}
where $\mathbf{c} = [C_1, C_2, \ldots, C_K] $ is the common rate vector.
In order to achieve the maximum sum-rate for the system, the BS should be able to allocate the power to the common and private messages in the way that the total power does not exceed the power of the BS. 
Given the power allocation coefficient vector $\pmb{\mu}$ and common rate vector $\mathbf{c}$, the optimization problem is then defined as follows:
\vspace{-0.3cm}
\begin{subequations}
\begin{align}
\max_{\pmb{\mu, c}} \quad & R_{sum} (\pmb{\mu,c })\\
\textrm{s.t.} \quad & \mu_c + \sum_{ k=1}^{K} \mu_k \leq 1,\\
\quad & \sum_{k=1}^{K}C_k \leq R_c, \\
\quad & C_k + R_k \geq Q_k, k \in \mathcal{K},\\
\quad & \mathbf{c} \geq 0.
\end{align}
\label{eq:max-sum-rate}
\end{subequations}
Constraint (\ref{eq:max-sum-rate}b) ensures that the sum of allocated power does not exceed the total power at the BS. Constraint (\ref{eq:max-sum-rate}c) guarantees that the common message can be decoded by all the users. Constraint (\ref{eq:max-sum-rate}d) is the minimum rate requirement (QoS) of user $u_k$. The final constraint (\ref{eq:max-sum-rate}e) is to guarantee the positive rate of the common message.

Optimizing (\ref{eq:max-sum-rate}) is challenging under the dynamic and uncertainty of the channel as the channel gain $\mathbf{h}_k$ between the BS and user $u_k$ varies over time, and channel state distribution is unknown by the BS. Unlike conventional multiple access schemes, splitting messages into different parts makes the problem even more challenging because the power needs to be allocated in the way that all the messages are decodable.
To model the dynamic of the channel state, we first formulate the problem by using the MDP framework. 

\vspace{-0.4cm}
\section{Problem Formulation}
To formulate the problem by using the MDP, we define a tuple $(\mathcal{S}, \mathcal{A}, \mathcal{P}, r, \gamma)$, where $\mathcal{S}$ is the state space, $\mathcal{A}$ is the action space, $\mathcal{P}: \mathcal{S} \times \mathcal{A} \times \mathcal{S} \rightarrow \mathbb{R}$ is the state transition probability distribution, $r:\ \mathcal{S} \times \mathcal{A} \rightarrow \mathbb{R}$ is the reward function, and $\tau \in (0, 1)$ is the discount factor.
\vspace{-0.45cm}
\subsection{State Space and Action Space}
\label{sec:action-space}
The state space of the BS is defined as:
$\mathcal{S} = \Big \{\gamma_k^c, \gamma_k^p \}; 1 \leq k \leq K \Big\}$, 
where $\gamma_k^c$ and $\gamma_k^p$ are the SINR feedbacks of the common and private messages from user $u_k$, respectively. 
The SINR feedbacks contain estimation errors due to the imperfect channel state information.
The action space of the BS is defined as: $\mathcal{A} = \{\pmb{\mu, c} \}$,
where $\pmb{\mu}$ and $\mathbf{c}$ are the power allocation coefficient vector and common rate vector, respectively. 

\vspace{-0.4cm}
\subsection{Reward Function}
The reward function is designed to maximize the sum-rate of the BS as in (\ref{eq:max-sum-rate}). 
To encourage the BS to optimize the sum-rate while all the QoS requirements of users are taken into account, we penalize the BS for each violated user's QoS. 
At current time step $t$, the BS observes the current state $s_t \in \mathcal{S}$, takes action $a_t\in \mathcal{A}$, and receives an immediate reward $r_t(s_t, a_t)$. The immediate reward can be defined as follows:
\vspace{-0.2cm}
\begin{equation}
\label{eq:reward}
r_t(s_t, a_t) = R_{sum}(1 - p_t),
\vspace{-0.1cm}
\end{equation} 
where $p_t$ is the penalty received by the BS for action $a_t$ that does not satisfy the QoS constraint in (7).
In particular, the penalty $p_t$ at the time step $t$ can be defined as follows:
\vspace{-0.2cm}
\begin{equation}
\label{eq:penalty}
p_t = \frac{1}{K}\sum_{k=1}^{K} \chi(C_k + R_k - Q_k),
\vspace{-0.1cm}
\end{equation}
where the function $\chi(C_k + R_k - Q_k)$ is equal to 1 if $C_k + R_k - Q_k < 0$, and otherwise $\chi(C_k + R_k - Q_k)=0$. If all the users' QoS are guaranteed, we have $p_t = 0$ and $r_t(s_t, a_t) = R_{sum}$. If none of the users' QoS is guaranteed, we have $p_t = 1$ and $r_t(s_t, a_t) = 0$.
Unlike the QoS constraint, the power and common rate constraints must not be violated at any given time step because the total transmission power at the BS is limited and the common message must be decodable at all users. Therefore, we do not include the penalties for the power and common rate constraints in the immediate reward. Alternatively, these constraints are treated as a part of our algorithm design, which is further discussed in Section~\ref{sec:ppo-algo}.

\vspace{-0.4cm}
\subsection{Optimization Formulation}

Let $\psi$ denote a stochastic policy (i.e., $\psi: \mathcal{S} \times \mathcal{A} \rightarrow [0, 1]$) which is the probability that action $a_t$ is taken at time step $t$ given the state $s_t$, i.e., $\psi = \text{Pr}\{a_t|s_t\}$. Given the discount factor $\tau \in (0, 1)$, let $J(\psi)$ denote the expected discounted reward of the BS by following policy $\psi$:
\vspace{-0.2cm}
\begin{equation}
\vspace{-0.2cm}
\label{eq:discounted-reward}
J(\psi) = \mathbb{E}_{a_t \sim \psi, s_t \sim \mathcal{P}} \Big[\sum_{t=0}^{\infty}\tau^t r_t(s_t, a_t)\Big].
\end{equation}
Our goal is to find the optimal policy $\psi^*$ for the BS that maximizes $J(\psi$), i.e., 
\vspace{-0.2cm}
\begin{equation}
\label{eq:max-return}
\begin{aligned}
\max_{\psi} \quad & J(\psi)\\
\textrm{s.t.} \quad & a_t \sim \psi(a_t|s_t), s_{t+1} \sim \mathcal{P}(s_{t+1}|s_t, a_t).
\end{aligned}
\end{equation}
Note that the state transition probability distribution $\mathcal{P}(s_{t+1}|s_t, a_t)$ is unknown to the BS.
Maximizing $J(\psi)$ is very challenging as we consider that the state and action spaces are continuous. Thus, conventional (deep) reinforcement learning methods (e.g., Q-learning and DQN)  cannot be directly adopted.
In this paper, we propose to use the Proximal Policy Optimization (PPO) algorithm~\cite{schulman2017} to approximate the optimal policy of the BS. The PPO is a sample-efficient algorithm which can work under the large continuous state and action spaces and can deal with the uncertainty of the channel state. 

\vspace{-0.3cm}
\section{Proximal Policy Optimization Algorithm}
\label{sec:ppo-algo}
Because the policy of the continuous action space cannot be obtained by using a conventional action-value method (e.g. DQN), PPO uses a policy's parameter vector to efficiently update the policy. The parameter vector, denoted as $\pmb{\theta}$, can be a linear vector or a nonlinear function approximator (i.e., a deep neural network)~\cite{schulman2017}. As a result, the optimal policy can be approximated as $\psi^* \leftarrow \psi_{\pmb{\theta}}$ with $\psi_{\pmb{\theta}}(a_t|s_t) = \text{Pr}\{a_t|s_t;\pmb{\theta}\}$.
The parameter vector $\pmb{\theta}$ can be updated by using a gradient ascent method as follows:
\vspace{-0.2cm}
\begin{equation}
\vspace{-0.2cm}
\label{eq:theta-update}
\pmb{\theta}_{t+1} = \pmb{\theta}_t + \alpha \hat{g}_t,
\end{equation}
where $\alpha$ is the step size, and $\hat{g}_t$ is a gradient estimator. The gradient estimator $\hat{g}_t$ can be calculated by differentiating a loss function as follows:
\vspace{-0.2cm}
\begin{equation}
\vspace{-0.1cm}
\label{eq:gradient-estimator}
\hat{g}_t = \nabla_{\pmb{\theta}} L(\pmb{\theta}).
\end{equation}
 We can observe from (\ref{eq:theta-update}) and (\ref{eq:gradient-estimator}) that the choice of the loss function $L(\pmb{\theta})$ has significant impact on the policy update. $L(\pmb{\theta})$ should have a small variance so that it does not cause bad gradient updates which result in significant decreases of $J(\psi)$. Since continuous action space is sensitive to the policy update, a minor negative change in updating $\pmb{\theta}$ can lead to destructively large policy updates~\cite{schulman2017}.
To overcome this problem, PPO algorithm uses a loss function $L^{PPO}(\pmb{\theta})$ to replace $L(\pmb{\theta})$:
\vspace{-0.2cm}
\begin{equation}
\label{eq:loss-ppo}
 L^{PPO}(\pmb{\theta}) = \min \Big(\frac{\psi_{\pmb{\theta}}}{\psi_{\pmb{\theta}_{old}}} A^{\psi_{\pmb{\theta}}}, u(\epsilon, A^{\psi_{\pmb{\theta}}}) \Big),
\end{equation}
where $A^{\psi_{\pmb{\theta}}}$ is the advantage function and $u(\epsilon, A^{\psi_{\pmb{\theta}}})$ is the clip function. 
The advantage function measures whether or not the action is better or worse than the policy's default behavior.
The clip function guarantees the policy does not change significantly after each update.

The advantage function at time step $t$ can be defined by:
\begin{equation}
A_t^{\psi_{\pmb{\theta}}}(s_t, a_t;\pmb{\theta}) = Q_t^{\psi_{\pmb{\theta}}}(s_t, a_t;\pmb{\theta}) - V_t^{\psi_{\pmb{\theta}}}(s_t;\pmb{\theta}),
\end{equation}
where $Q_t^{\psi_{\pmb{\theta}}}(s_t, a_t;\pmb{\theta}) = \mathbb{E}_{a_t \sim \psi_{\pmb{\theta}}, s_t \sim \mathcal{P}}\Big[\sum_{l=0}^{\infty}\tau^l r_t(s_{t+l}, a_{t+l})\Big]$ is the action value function and $V_t^{\psi_{\pmb{\theta}}}(s_t;\pmb{\theta}) = \mathbb{E}_{s_t \sim \mathcal{P}}\Big[\sum_{l=0}^{\infty}\tau^l r_t(s_{t+l}, a_{t+l})\Big]$ is the state value function. 
The clip function is thus defined as follows:
\vspace{-0.2cm}
\begin{equation}
u(\epsilon, A^{\psi_{\pmb{\theta}}}) = 
\begin{cases}
      (1+\epsilon)A^{\psi_{\pmb{\theta}}}, & \text{if}\ A^{\psi_{\pmb{\theta}}} \geq 0, \\
      (1-\epsilon)A^{\psi_{\pmb{\theta}}}, & \text{if}\ A^{\psi_{\pmb{\theta}}} < 0.
    \end{cases}
\end{equation}

The idea of PPO is to prevent the new policy from being attracted to go far away from the old policy $\psi_{\pmb{\theta}_{old}}$. 
The first term inside the $\min$ operator in (\ref{eq:loss-ppo}), i.e., $\frac{\psi_{\pmb{\theta}}}{\psi_{\pmb{\theta}_{old}}} A^{\psi_{\pmb{\theta}}}$, is the surrogate objective which takes into consideration the probability ratio between the new policy and old policy, i.e., $\frac{\psi_{\pmb{\theta}}}{\psi_{\pmb{\theta}_{old}}}$. The second term, i.e., $u(\epsilon, A^{\psi_{\pmb{\theta}}})$, removes the incentive for moving this probability ratio outside of the interval $[1-\epsilon, 1+\epsilon]$.

In this paper, we use a deep neural network as a nonlinear function approximator to approximate the policy $\psi_{\pmb{\theta}}$ and advantage function $A^{\psi_{\pmb{\theta}}}$. The input of the network is the state of the environment, i.e., $s_t = \{\gamma_k^c, \gamma_k^p; \forall k \in \mathcal{K}\}$. The output is the joint power allocation and common rate vector $a_t = [\mu_c, \mu_1, \ldots, \mu_K, C_1, \ldots, C_K]$. To ensure that the power constraint in (\ref{eq:max-sum-rate}b) and the common rate vector constraint in (\ref{eq:max-sum-rate}c), we use the Softmax activation function for the output layer of the network so that $\mu_c + \sum_{ k = 1}^{K} \mu_k = 1$, and $\sum_{k=1}^{K} C_k = R_c$.

\vspace{-0.3cm}
\section{Performance Evaluation}
\subsection{Parameter Settings}
We consider the total transmission power of the BS to be $P_t=40$ (dBm).
The number of antennas of the BS and the number of users are set as $M = K = 4$. The channel estimation $\mathbf{h}_k$ at the BS contains estimation error, i.e., $\mathbf{h}_k = \mathbf{\hat{h}}_k + \mathbf{\tilde{h}}_k$, where $\mathbf{\hat{h}}_k$ is the actual channel, $\mathbf{\tilde{h}}_k$ is the channel estimation error.
The mean value of $\mathbf{\tilde{h}}_k$ is inversely proportional to the transmission power, i.e., $ \mathbb{E}\big\{||\mathbf{\tilde{h}}_k|| ^2\big\} \sim P_t^{-0.6}$~\cite{yijie2018}. 
 The QoS requirements are assumed to be the same at each user, i.e., $Q_k = Q_m = 0.1$(bps/Hz). 

We first evaluate the performance of the proposed PPO algorithm with two baseline schemes that are Q-learning and Greedy. Because Q-learning is an action-value method which cannot be directly applied for the continuous state/action problem, we discretize the state and action spaces as follows.
In Q-learning, we divide each dimension of the state space into two levels. To discretize the action space, we adopt an uniform power allocation mechanism~\cite{dai2016}. We consider 9 discrete actions of the Q-learning and 99 discrete actions of the Greedy scheme. As a result,  Q-learning algorithm maintains a Q-table of $2^{10} \times 9$ Q-values.
With Greedy algorithm, all historical reward values are stored in the memory and the BS keeps selecting the action that obtains the highest reward, compared to the historical rewards. 
Otherwise, the BS randomly selects other actions to further explore the environment. 
It is noted that the state/action space quantization above is applied for the baseline schemes and the proposed PPO algorithm still considers the complete continuous state and action spaces.

\vspace{-0.4cm}
\subsection{Simulation Results}
%\vspace{-0.1cm}
\begin{figure}[t]
\centering
\includegraphics[width=0.54\linewidth]{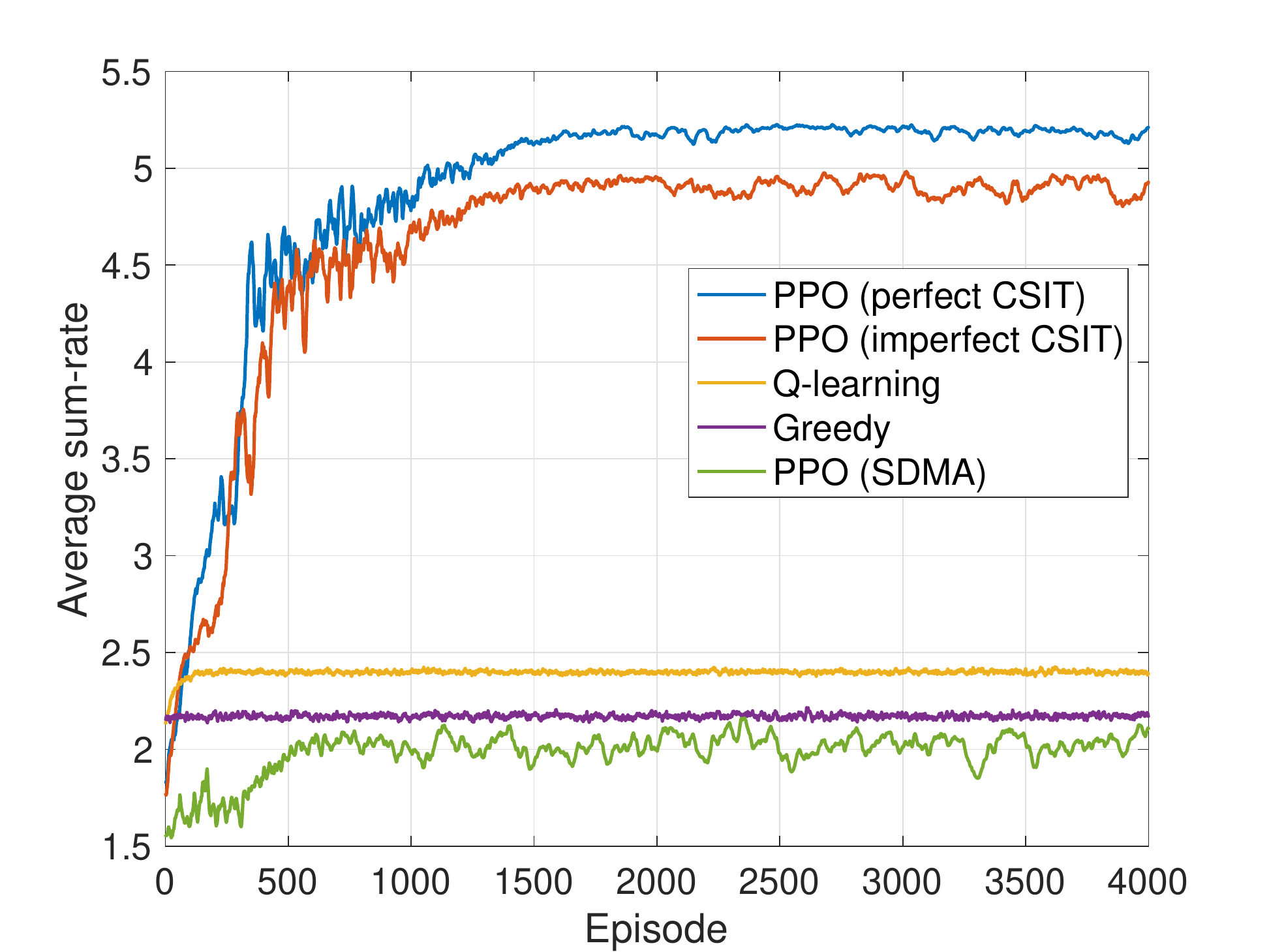}
\caption{Learning curves of the algorithms with $P_t = 40$ (dBm) and $Q_m = 0.1$ (bps/Hz).}
\vspace{-0.3cm}
\label{fig:learning-curves}
\end{figure}
\vspace{-0.1cm}
\begin{figure}[t]
\centering
	\begin{subfigure}[b]{0.246\textwidth}
	 	\centering
	 	\includegraphics[width=\textwidth]{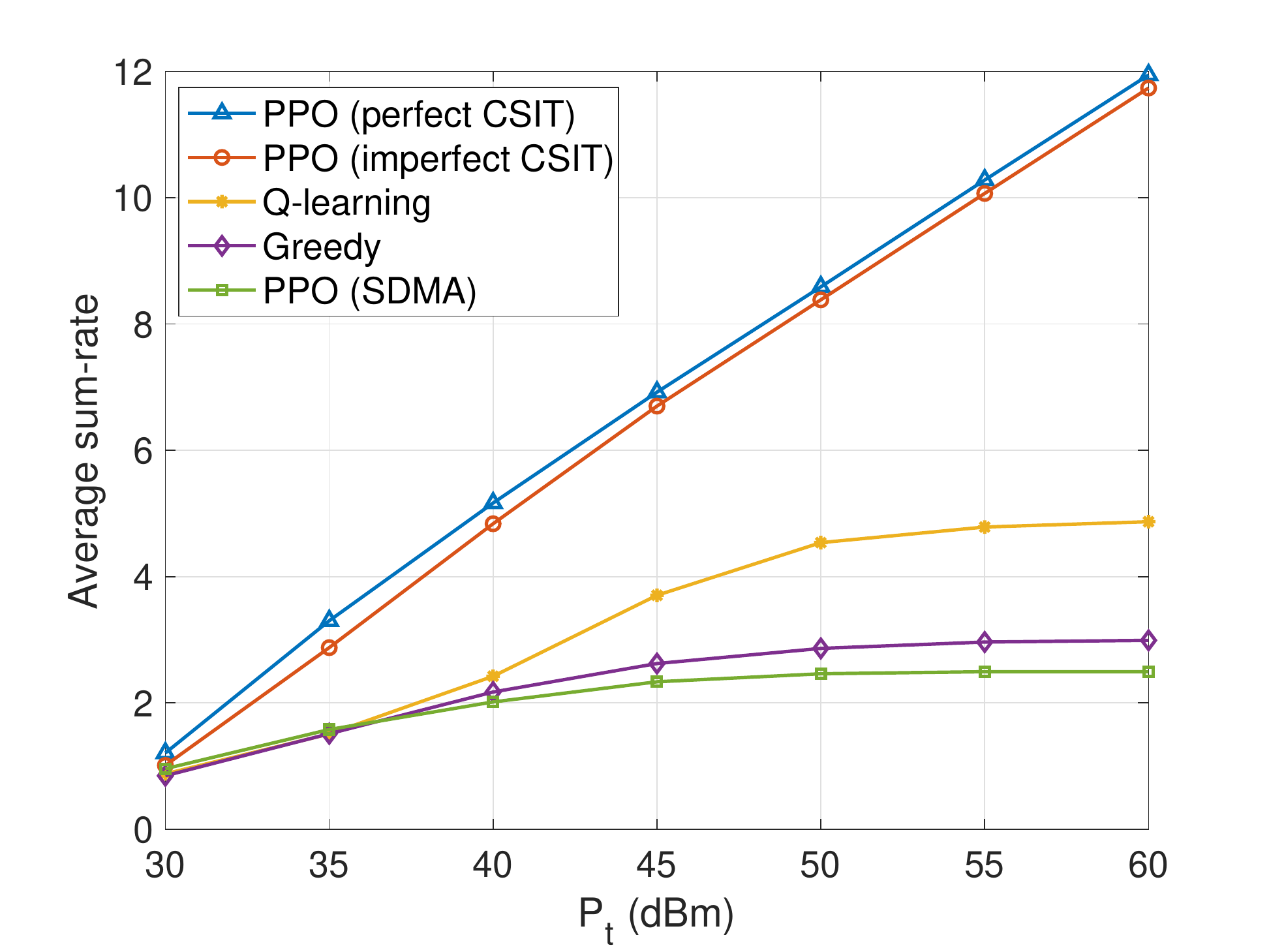}
	 	\vspace{-0.5cm}
	 	\caption{}
	 \end{subfigure}
	 \hspace{-0.6cm}
	\begin{subfigure}[b]{0.25\textwidth}
	 	\centering
	 	\includegraphics[width=\textwidth]{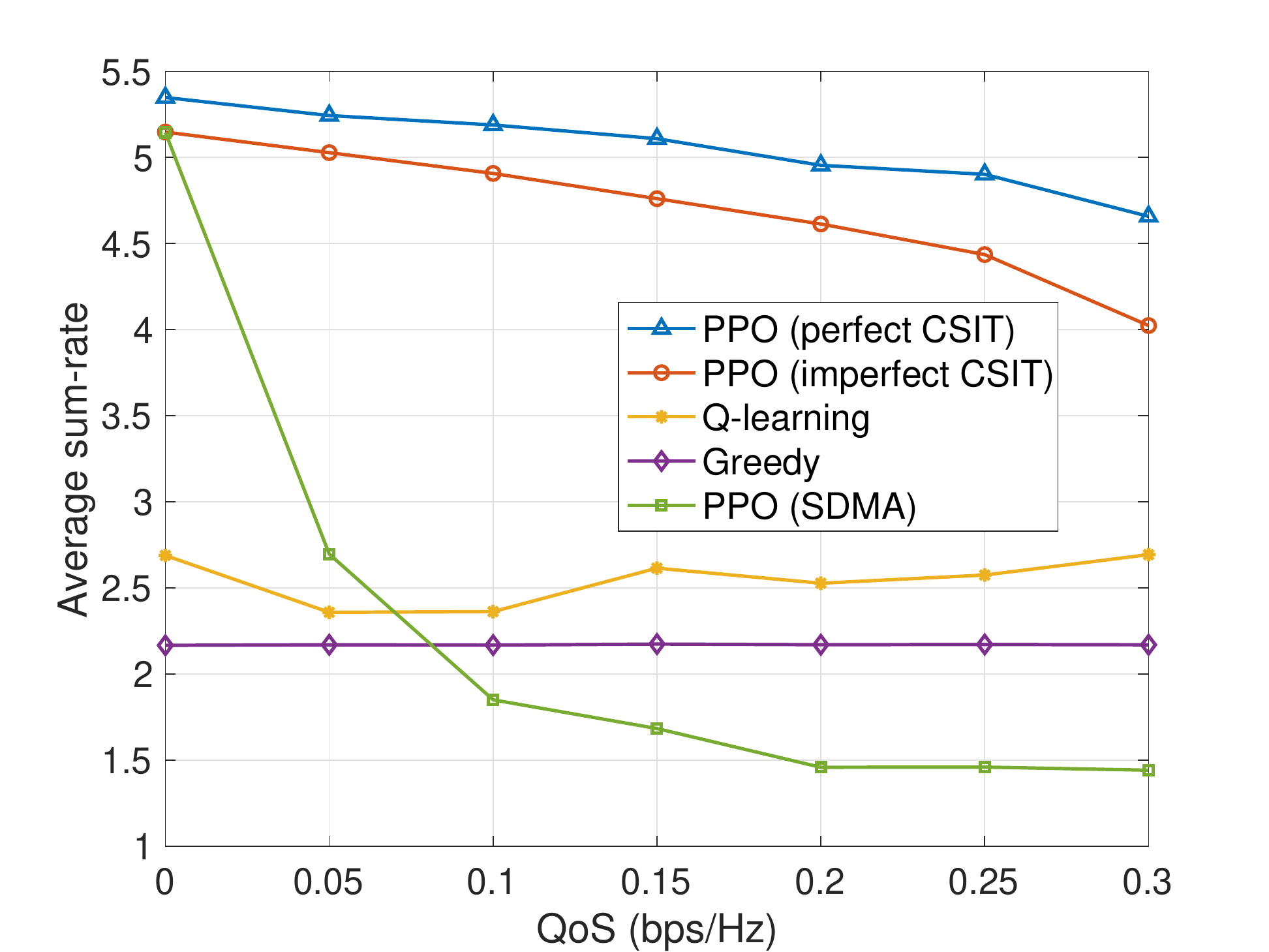}
	 	\vspace{-0.5cm}
	 	\caption{}
	 \end{subfigure} 
\caption{(a) Average sum-rate vs. total transmission power at the BS ($P_t$) and (b) average sum-rate vs. minimum QoS requirement ($Q_m$).}
\vspace{-0.3cm}
\label{fig:power-varies}
\end{figure}

In Fig.~\ref{fig:learning-curves}, we show the learning curves of the three algorithms in the first 4,000 episodes in which each epsiode has the length of 200 time steps. To further evaluate the advantages of RSMA over conventional techniques, we use SDMA as a baseline scheme. For a fair comparison, we do not include NOMA in the simulation since NOMA requires a more complex architecture, i.e., multiple layers of SIC, for decoding the messages.
We also evaluate the proposed PPO algorithm for RSMA in both perfect and imperfect CSIT scenarios.

The proposed PPO algorithm clearly outperforms the baseline schemes in terms of average sum-rate. The reason is that with the large number of states and actions, Q-learning is unable to update all the Q-values to obtain the desired optimal policy. This is also known as the curse-of-dimensionality problem. Furthermore, the state/action space quantization may also remove states and actions that are valuable in the policy update process. For the Greedy scheme, the connection between the state, action and the policy is not considered, which yields a much lower performance. The sum-rate obtained by SDMA with PPO is  much lower than those of RSMA.

Next, we vary the transmission power capacity at the BS and evaluate the performance of the three algorithms as shown in Fig.~\ref{fig:power-varies}(a).
% Each point in Fig.~\ref{fig:power-varies} is obtained by averaging values of 500 episodes from episode 1500 to episode 2000 when the algorithms converged.
As the transmission power increases, the average sum-rate obtained by all algorithms increase, and the proposed PPO always achieves the best performance compared to those of the Q-learning and Greedy. In particular, when the transmission power of the BS is 60 dBm, the average sum-rate obtained by the proposed PPO algorithm can achieve up to 11.9 and 11.7 with perfect and imperfect CSIT, respectively, which are significantly greater than those of the Q-learning and Greedy (i.e., 4.7 and 2.9). Similar to the results obtained in Fig.~\ref{fig:learning-curves}, the sum-rate values obtained by SDMA are much lower than those of RSMA with all transmission power values.

%We can observe that the proposed PPO algorithm outperforms the baselines in terms of average reward (Fig.~\ref{fig:power-varies}(a)) and average sum-rate (Fig.~\ref{fig:power-varies}(b)). The three algorithms increase their average reward and average sum-rate values when the power at the BS increases from 5 to 40 dBm. The reason is that the reward is proportional to the sum-rate as defined in (\ref{eq:reward}). Therefore, a higher reward results in a higher sum-rate. In particular, the PPO algorithm can increase sum-rate up to 563\%, compared to the Greedy scheme. 

Finally, we vary the QoS requirements to evaluate the performance of the three algorithms as shown in Fig.~\ref{fig:power-varies}(b).
As the QoS requirements increase, the average sum-rate obtained by all the algorithms decrease. The reason for this is that at high rate requirements, the BS cannot satisfy the constraints of all the users and thus it is penalized by the penalty $p_t$ as defined in (\ref{eq:reward}). However, our proposed PPO algorithm still achieves the best performance given all the QoS requirements. 

\vspace{-0.25cm}
\section{Conclusion}
%\vspace{-0.1cm}
In this letter, we have developed a highly effective framework to maximize the sum-rate for RSMA networks under the dynamic and uncertainty of the communication channel. Specifically, we have first formulated the problem with MDP framework and then proposed a deep reinforcement learning algorithm to quickly find the optimal power allocation policy. Our proposed method does not require any information of the channel state in advance and can deal with the continuous state and action spaces.
Simulation results have shown that our proposed scheme can outperform baseline schemes in terms of average sum-rate under different power and QoS constraints.


\vspace{-0.3cm}
\begin{thebibliography}{100}
\bibliographystyle{IEEEtranS}
\vspace{-0.15cm}

\bibitem{dizdar2020}
O. Dizdar, \textit{et al.}, ``Rate-splitting multiple access: A new frontier for the PHY layer of 6G," \textit{arXiv preprint} arXiv:2006.01437, 2020.

\bibitem{dai2016}
M. Dai, \textit{et al.}, ``A rate splitting strategy for massive MIMO with imperfect CSIT," \textit{IEEE Trans. Wireless Commun.}, vol. 15, no. 7, pp. 4611-4624, Mar. 2016.

\bibitem{joudeh2016}
H. Joudeh and B. Clerckx, ``Robust transmission in downlink multiuser
MISO systems: a rate-splitting approach,” \textit{IEEE Trans. Signal Process.},
vol. 64, no. 23, pp. 6227-6242, Dec. 2016.

\bibitem{li2020}
P. Li, \textit{et al},  ``Cooperative rate-splitting for secrecy sum-rate enhancement in multi-antenna broadcast channels," in \textit{IEEE 31st Annu. Int. Symp. on Pers., Indoor and Mobile Radio Commun.}, 2020.

\bibitem{joudeh2016sumrate}
H. Joudeh and B. Clerckx, ``Sum-rate maximization for linearly precoded downlink multiuser MISO systems with partial CSIT: A rate-splitting approach, " \textit{IEEE Trans. Commun}, vol. 64, no. 11 , pp. 4847-4861, Aug. 2016.

\bibitem{he2019}
C. He, \textit{et al.}, ``Joint power allocation and channel assignment for NOMA with deep reinforcement learning," \textit{IEEE J. Sel. Areas Commun.}, vol. 37, no.10, pp. 2200-2210., Aug. 2019.

\bibitem{schulman2017}
J. Schulman, \textit{et al.}, ``Proximal policy optimization algorithms," \textit{arXiv preprint}, arXiv:1707.06347, 2017.

\bibitem{yijie2018}
Y. Mao, B. Clerckx, and V. O. Li, ``Rate-splitting multiple access for downlink communication systems: bridging, generalizing, and outperforming SDMA and NOMA," \textit{EURASIP J. Wireless Commun. Netw.}, no. 1, pp. 1-54, Dec. 2018.

\end{thebibliography}
\end{document}